\documentclass[twocolumn,showpacs,pre,aps]{revtex4}
\usepackage{dcolumn}
\usepackage{bm}

\begin{document}

\title{Excitation of magnetostatic spin waves in
ferromagnetic films}

\author{Yu. Kuzovlev, N. Mezin and G. Yarosh}
\email{kuzovlev@kinetic.ac.donetsk.ua} \affiliation{A.A.Galkin
Physics and Technology Institute of NASU, ul. R.Luxemburg 72,
83114 Donetsk, Ukraine}

\date{\today}

\begin{abstract}
The torque equation of nonlinear spin dynamics is considered in
the magnetostatic approximation. In this framework, exact
expressions for propagator of linear magnetostatic waves in
ferromagnetic film between two antennas and corresponding mutual
impedance of the antennas are derived, under conditions of uniform
but arbitrarily oriented static magnetization and arbitrary
anisotropy. The results imply also full description of spectrum of
the waves.
\end{abstract}


\maketitle


\section{Introduction}
The terms ``magnetostatic oscillations'' or ``magnetostatic
waves'' (MSW) \cite{Walker,de,White,ABP} mean such relatively long
spin waves (SW) whose properties are dominated by the long-range
quasi-static dipole interaction between ``spins'' and eventually
by geometry of ferromagnetic sample (its shape and dimensions). In
opposite, properties of relatively short SW (spin waves in narrow
sense) \cite{White,ABP,LP} come rather from the short-range
exchange interaction. In ferrite films, MSW can be easy excited,
with the help of microwave-frequency magnetic fields induced by
wire or strip-like antennas \cite{wig,kalin3,kalin4}, up to the
level of strongly nonlinear interactions of MSW between themselves
and with SW. In addition, small velocity of MSW ensures
compactness of nonlinear transformations of electromagnetic
signals. When exploring these possibilities, it is important to
know characteristics of linear MSW modes and corresponding linear
impedances of their antennas. Concerning MSW in flat homogeneous
films, there are widely used exact results by Damon and Eshbach
\cite{de}. Other authors \cite{kalin1,kalin2} (see also paper by
Kalinikos and coworkers in \cite{wig} and references therein)
developed approximate approaches to films which are
non-homogeneous because of specific surface effects. At the same
time, to the best of our knowledge, the homogeneous case remains
incompletely investigated. In the present paper we find new
solutions to this case believing that they will be useful for both
practice and theoretical modeling of more complicated situations.

We confine our consideration by ``magnetostatic approximation''
which neglects SW (in narrow sense) at all, as if radius of the
exchange interaction, $r_e\,$, is equal to zero. This formal trick
seems reasonable when cross-section dimensions of all the antennas
or/and distances from them to film's surface are much greater than
$r_e\,$. Indeed, under this condition magnetic field created by
antennas is so smoothly distributed in film's interior (with
spatial scales much greater than $r_e$) that must only very poorly
excite those SW whose length is comparable with $r_e\,$ or
smaller. We do not know strict proof of this statement, but it
rather convincingly follows from results of the ``magnetostatic
approximation'' itself (in particular, results presented below)
which demonstrate, at mentioned condition, insignificance of
``infinitely short'' MSW in the sense of both their private
excitations and summary contribution to film's linear response to
weak external field. Other deal is when in definite nonlinear
processes even SW much shorter than $r_e$ can be generated from
long MSW \cite{wig} (by the way, all the nonlinear models
collected in \cite{wig} agree with the above statement). However,
here we are interested in linear (small-amplitude) response only,
although examining the path to it from basic nonlinear equations.

Besides, firstly, we will keep in mind that film under
consideration is thick, in the sense that its thickness, $D\,$, is
large as compared with $r_e\,$. This condition will make our
results be better applicable to real films with surface-induced
non-uniformity. Secondly, if a ground (static) state of film's
magnetization is divided into domains and hence strongly
non-uniform then this state itself must be understood and
described before its magnetostatic excitations. Of course, we
avoid this extremely difficult problem, being satisfied by
consideration of MSW on background of uniformly magnetized static
state. Fortunately, for example, in yttrium-iron garnet films such
state can be enforced by comparatively small bias field. At
formulated assumptions, we will obtain the function (``propagator
of MSW'') which describes mutual influence of two antennas by
means of MSW propagating in a film with arbitrarily directed
ground magnetization in presence of arbitrary magnetic anisotropy.

\section{Basic equations}
The classical semi-phenomenological dynamics of magnetization in
solid ferromagnets and ferrites \cite{ABP,LP,LL} is based on the
Landau-Lifshitz-Gilbert equation (the torque equation)
\begin{equation}
\frac{d\bm{S}}{dt}=\bm{F}\times\bm{S}+ \gamma\{{\bm{F}-\bm{S}
(\bm{S}\cdot\bm{F})\}}  \label{eq:LL}
\end{equation}
Here $\bm{S}\,$ is unit vector in the direction of local
magnetization ($|\bm{S}|=1$); $\bm{F}$ is thermodynamic force, or
effective internal magnetic field, expressed in units of the
saturation magnetization, $M_s$; the time is expressed in units of
$\tau_0=$ $(2\pi gM_s)^{-1}\,$ ($g\approx 2.8\,$MHz/Oe is the
giromagnetic ratio); $\,\times\,$ and $\,\cdot\,$ are symbols of
vector and scalar products, respectively; $\,\gamma\,$ is
phenomenological friction (dissipation) parameter. For our
purposes, the simplest model of dissipation is sufficient as those
introduced by Eq.\ref{eq:LL}.

The internal field, $\bm{F}$, is composed at least from
\cite{ABP,LP,LL} (i) external bias field, $\bm{H}_e\,$, (ii)
magnetic anisotropy field, $\bm{H}_a\,$, (iii) own magnetic field
created by the magnetization (by ``spins''), $\bm{H}_s\,$, and
(iv) exchange field, $\bm{H}_{exch}\,$. We do not add to this list
magnetic field induced by electric currents in the ferromagnetic
sample since assume that it is good isolator. Usually small
dimensions of real samples allow to neglect time delay of
$\bm{H}_s\,$, therefore $\bm{H}_s\,$ can be found from
quasi-static version of the Maxwell equations. Particularly, in
absence of conductors and other ferromagnets in vicinity of the
sample, these equations yield
\begin{equation}
\bm{H}_s=-\,\widehat{G}\bm{S}\equiv
\bm{\nabla}\left(\bm{\nabla}\cdot \int \frac {\bm{S}(r\prime
)dr\prime}{|r-r\prime |}\right) \label{eq:G}
\end{equation}
In most simple model of exchange interaction, $\,\bm{H}_{exch}=$
$r_e^2\nabla ^2\bm{S}\,$ \cite{ABP,LP,LL}. It should be noted that
exchange interaction ensures local smoothness of the magnetization
distribution, $\bm{S}\,$, while the operator of dipole-dipole
interaction, $\widehat{G}\,$, is bounded when acts on smooth
distributions: $||\widehat{G}||\leq 4\pi\,$. If $A(\bm{S})$
denotes density of energy of the anisotropy then
\[\bm{H}_a=-A^{\prime }(\bm{S})\equiv -\,\partial
A(\bm{S})/\partial \bm{S}
\]
In principle, all the consideration in this section and next
section can be easily generalized to non-uniform anisotropy if
suppose that $A(\bm{S})$ is some function of space coordinates.

Let the subscript ``0'' be attribute of static magnetization state
at constant bias field,  $\bm{H}_e=$ $\bm{H}_0=$ $const$.
According to Eq.\ref{eq:LL} in any such state the vectors
$\bm{S}_0$ and $\bm{F}_0$ are parallel one to another, that is
$\bm{F}_0=W_0\bm{S}_0$, where scalar field $\,W_0\,$ (absolute
value of static internal magnetic field) is defined by the
requirement $|\bm{S}|=1$.

Let additional time-varying magnetic field is switched on:
$\bm{H}_e=$ $\bm{H}_0+$ $\bm{h}\,$, $\bm{h}=$ $\bm{h}(t)\,$.
Corresponding deviation of magnetization from its static value,
$\bm{s}=$ $\bm{S}-\bm{S}_0\,$, can be represented as
\begin{equation}
\bm{s}=\bm{S}_{\bot }+(S_{\Vert }-1)\bm{S}_0\,\,,\,\,\,
\bm{S}_{\bot }=\widehat{\Pi}\bm{S}\,\perp \bm{S}_0\,\,,
\label{eq:s}
\end{equation}
\[
\widehat{\Pi}\equiv 1-\bm{S}_0\otimes \bm{S}_0\,\,,\,\,\,\,\,\,
S_{\Vert }=\pm\sqrt{1-|\bm{S}_{\bot }|^2}\,\,,
\]
where $\otimes $ means tensor product, and $\widehat{\Pi}$ is
operator (matrix) which performs projection of vectors onto the
plane perpendicular to $\bm{S}_0$. If none spins are overturned by
the perturbation then sign of $S_{\Vert }$ is everywhere positive.
In terms of $\,\bm{s}\,$ and $\bm{S}_{\bot }\,$, Eq.1 transforms
into
\begin{equation}
\frac{d\bm{S}_{\bot }}{dt}=S_{\Vert }[\bm{F}_{\bot
}\times\bm{S}_0]+ \gamma (1-\bm{S}_{\bot}\otimes \bm{S}_{\bot
})\bm{F}_{\bot }\,\,,  \label{eq:Eq}
\end{equation}
where $\,\bm{F}_{\bot }=-\,\delta E_{\bot}/\delta \bm{S}_{\bot }$,
and $\,E_{\bot}\,$ is the excess energy (energy of excitation)
implied by the perturbation,
\begin{equation}
E_{\bot }=\int \left [\frac {1}{2} W_0\bm{s}^2+
\widetilde{A}(\bm{s})+\frac{1}{2}\bm{s}\cdot\widehat{G}\bm{s}-
\bm{h}\cdot\bm{s}+C(\bm{\nabla}\bm{s}) \right ]dr \label{eq:En}
\end{equation}
Here function $\widetilde{A}\,$ is defined by
\[
\widetilde{A}(\bm{s})=A(\bm{S}_0+\bm{s})-A(\bm{S}_{0})-
\bm{s}\cdot A^{\prime}(\bm{S}_{0})\,\,,
\]
and function $C\,$ represents exchange contribution to the excess
energy. In the mentioned model, $C(\bm{\nabla}\bm{s})=$ $\frac
{1}{2}r_e^2\sum_{\alpha\beta} (\nabla_{\alpha}s_{\beta})^2$. The
functional derivative $\bm{F}_{\bot }$ in Eq.\ref{eq:Eq} should be
evaluated with taking into account full dependence of $\,\bm{s}\,$
on $\,\bm{S}_{\bot }\,$ in accordance with Eq.\ref{eq:s}.

Importantly, the frictionless version of Eq.\ref{eq:Eq} (that is
at $\gamma =0$) follows from the variational principle
\begin{equation}
\delta \int \left\{ \int \left(\bm{S}_0\cdot \left[
\frac{d\bm{S}_{\bot }}{dt}\times \bm{S}_{\bot }\right] \right)
\frac{dr}{1+S_{\Vert }}+ E_{\bot}\right\} dt=0 \label{eq:vp}
\end{equation}

Of course, in general $\bm{S}_0$ is a complicated function of
spatial coordinates, hence all the related values ($W_0$,
$\widehat{\Pi}$, $\widetilde{A}(\bm{s})$, and so on) are space
dependent.

\section{Linear waves}
If the functional (\ref{eq:En}) represents positively defined
quadratic form then the static magnetization pattern $\bm{S}_0$ is
stable with respect to any small perturbation, and therefore we
can speak about linear eigenmodes of the excitation. In this case,
let us introduce the spin precession operator, $\widehat{R}\,$,
the anisotropy matrix, $\widehat{A}\,$, and besides the exchange
operator, $\widehat{C}\,$, by the relations
\[
\widehat{R}\bm{V}\equiv \bm{S}_0\times \bm{V}\,\,,\,\,\,\,
\widehat{C}\bm{V}\equiv -\,r_e^2\nabla ^2\bm{V}\,\,,
\]
\[
\widehat{A}_{\alpha \beta}=\partial ^2A(\bm{S}_0)/\partial
\bm{S}_{0\alpha}\partial \bm{S}_{0\beta}
\]
For the linear regime Eq.\ref{eq:Eq} yields
\begin{equation}
\frac{d\bm{S}_{\bot }}{dt}=(\widehat{R}-\gamma \widehat{\Pi
})(\widehat{W}\bm{S}_{\bot}-\bm{h})\,\,,\label{eq:sa}
\end{equation}
where we introduce the integral-differential operator
\[
\widehat{W}\equiv W_0+\widehat{A}+\widehat{G}+\widehat{C}
\]

Rejecting from Eq.\ref{eq:sa} both dissipation and external pump,
we obtain equations for SW and MSW eigenmodes and
eigenfrequencies:
\begin{equation}
\bm{S}_{\bot }\equiv \bm{V}e^{-i\omega t}\,\,,\,\,\,\, -i\omega
\bm{V}=\widehat{R}\widehat{W}\bm{V}       \label{eq:ew}
\end{equation}
Herewith it is sufficient to consider positive frequencies only.
Let the eigenmodes be enumerated by an index $k\,$. Since in a
stable state the operator $\widehat{W}$ is positively defined, we
can write
\begin{equation}
\omega _{k}\widetilde{\bm{V}}_{k}=i\widehat{W}^{1/2}\widehat{R}
\widehat{W}^{1/2}\widetilde{\bm{V}}_{k}\,\,,\,\,\,\,
\widetilde{\bm{V}}_{k}\equiv \widehat{W}^{1/2}\bm{V}_{k}
\label{eq:sc}
\end{equation}
The operator on right-hand side in the left of these two
equalities must be self-adjoint, hence, its eigenfunctions
$\widetilde{\bm{V}}_{k}$ can be made mutually orthogonal. From
here the orthogonality rule for the eigenmodes does follow:
\begin{equation}
i\int \bm{S}_{0}\cdot[\bm{V}_{k}\times \bm{V}_{m}^{\ast }]
\,dr=\delta _{mk}   \label{eq:ort}
\end{equation}
The same rule is specified by the variational principle
(\ref{eq:vp}).

To get more general formulation of linear theory, we should return
from the ``ready'' dipole interaction operator $\widehat{G}\,$ to
Maxwell equations:
\begin{equation}
\frac{d\bm{S}_{\bot }}{dt}=(\widehat{R}- \gamma
\widehat{\Pi})\{(W_0+\widehat{A}+\widehat{C})\bm{S}_{\bot}-
\bm{h}-\bm{h}_S\}\,\,,\label{eq:lin}
\end{equation}
\begin{equation}
\bm{\nabla }\cdot (\bm{h}_S+4\pi \bm{S}_{\bot })=0\,\,,\,\,\,\,
\bm{\nabla }\times \bm{h}_S=0\,\,,               \label{eq:lmax}
\end{equation}
where we introduced new vector field, $\bm{h}_S$, which represents
time-varying part of magnetic field self-induced by magnetization
(i.e. the same as the whole field induced by $\bm{s}$). As before,
it is assumed that the sample is non-conducting.

Applying Fourier transform to Eq.\ref{eq:lin}, in the frequency
representation (frequency domain) from (\ref{eq:lin}) and
(\ref{eq:lmax}) we have
\begin{equation}
\bm{S}_{\bot }=\widehat{\chi }\,[\bm{h}+\bm{h}_S]\,\,,\,\,\,\,\,
\bm{\nabla }\cdot\widehat{\mu }\,[\bm{h}+\bm{h}_S]=0\,\,,
\label{eq:fr}
\end{equation}
\begin{equation}
\widehat{\chi }=\{i\omega +(\widehat{R}- \gamma \widehat{\Pi
})(W_0+\widehat{A}+\widehat{C})\}^{-1}(\widehat{R}- \gamma
\widehat{\Pi })\,,               \label{eq:hi}
\end{equation}
where $\,\,\widehat{\mu }\equiv 1+4\pi \widehat{\chi }\,$.
Obviously, because of presence of the differential operator
$\widehat{C}$ in denominator of the polarizability matrix
$\widehat{\chi }\,$ in (\ref{eq:hi}), in fact $\widehat{\chi }\,$
is an integral operator.

At this point we go to the ``magnetostatic approximation''
formulated and discussed in Sec.1. Concretely, we reject the
exchange operator $\widehat{C}$ from denominator of (\ref{eq:hi}).
Formally, this is equivalent to that the exchange radius $r_e\,$
turns into zero (of course, herewith we do not neglect exchange
interaction since it remains responsible for the magnetization
phenomenon itself). Strictly speaking, simultaneously the static
magnetization, i.e. the patterns $\bm{S}_0$ and $W_0$, also must
be treated in this limit. But this does not matter in the case of
uniform static magnetization which we will investigate below.

After that, $\widehat{\chi }\,$ turns into algebraic expression
becoming literally matrix, and the problem reduces to purely
differential equations for the field $\bm{h}_S$. Direct analytical
calculation gives very simple expression for the polarizability:
\begin{equation}
\widehat{\chi }=\frac{(\overline{W}_0+A_{1}+A_{2}) \widehat{\Pi
}-\widehat{A}_{\bot }+i\widetilde{\omega }
\widehat{R}}{(\overline{W}_0+A_{1})(\overline{W}_0+A_{2})-
\widetilde{\omega }^{2}}\,\,\,,                \label{eq:mhi}
\end{equation}
\[
\widehat{A}_{\bot }\equiv \widehat{\Pi }\widehat{A}
\widehat{\Pi}\,\,,\,\,\,\,\, \overline{W}_0\equiv W_0-i\gamma
\widetilde{\omega }\,,\,\,\,\, \widetilde{\omega }\equiv
\frac{\omega }{1+\gamma ^{2}}
\]
Here $A_1$ and $A_2$ are those two eigenvalues of matrix
$\,\widehat{A}_{\bot }\,$ which correspond to the pair of its
eigenvectors perpendicular to $\bm{S}_0$ and one to another:
$\,\widehat{A}_{\bot }\bm{a}_{1,2}$ $=$ $A_{1,2}\,\bm{a}_{1,2}\,$.
We enumerate them so that $\bm{S}_0\cdot[\bm{a}_1\times \bm{a}_2]$
$>0\,$. At practically interesting ferrite samples $\gamma
\lesssim $ $10^{-3}$, therefore $\gamma ^2$ plays no role.

\section{Propagator of magnetostatic waves in films}
Let the time-varying field $\bm{h}\,$ is induced by some
conductors which are placed outside the ferromagnetic sample and
carry a.c. currents $I_n\,$ distributed with densities
$I_n\bm{J}_n\,$ ($n=$ $1,2,...$). Then we can write
\begin{equation}
\bm{h}=\sum \bm{h}_mI_m\,,\,\,\,\bm{\nabla}\cdot
\bm{h}_n=0\,\,,\,\, \bm{\nabla}\times\,\bm{h}_n=\frac{4\pi
}{c}\bm{J}_n \label{eq:cur}
\end{equation}
Here $\bm{h}_n\,$ is magnetic field created by unit-value current
in  $n$-th conductor. The same function determines voltage
(e.m.f.), $\varepsilon _n\,$, induced in the $n$-th conductor by
time-varying magnetization of the sample:
\begin{equation}
\varepsilon _n=\int \left(\bm{h}_n\cdot
\frac{d\bm{s}}{dt}\right)dr             \label{eq:emf}
\end{equation}
The fields $\bm{h}_n$ as well as the self-induced field $\bm{h}_S$
can be represented in the potential form.

In the linear regime, when $\bm{s}\rightarrow \bm{S}_{\bot }$, the
response of the sample divides into sum of partial responses:
\[
\bm{h}_n=-\bm{\nabla} U_n\,,\,\,\, \bm{h}_S=\sum
\widehat{\bm{h}}_{sm}I_m\,,\,\,\,
\widehat{\bm{h}}_{sn}=-\bm{\nabla} \widehat{U}_{sn}\,
\]
where $\widehat{U}_{sn}$ is potential of the field induced by the
sample in response to influence by $n$-th conductor. After
obtaining $\bm{h}_S\,$ in company with $\bm{S}_{\bot }$ from
(\ref{eq:emf}) we will determine mutual impedances of the
conductors, $\widehat{Z}_{nm}$, caused by their interaction из
through the ferromagnetic:
 $\,\varepsilon _n=\sum\widehat{Z}_{nm}I_{m}\,$
(the hat marks convolution operators).

Now concretize the sample as plate (film) whose in-plane
dimensions much exceed its thickness, formally as infinite plate.
At sufficiently large bias field, $\bm{H}_0$, ferromagnetic plate
allows for stable state of uniform magnetization. In real
finite-size films, such the state is demagnetized at film's edges
only, in strip-like regions whose width is few of $D$ ($D$ is
thickness). This justifies the theory of plane MSW in infinite
uniformly magnetized film.

Let film is disposed in the region $-D/2<z<D/2$. Naturally, make
Fourier transform with respect to time and in-plane coordinates,
$x\,$ and $y\,$, marking transformed functions by tilde. Introduce
designations  $\,\bm{k}=$ $\{k_x,k_y\}\,$, $\,\bm{\nabla} =$
$\{ik_x,ik_y,$ $\nabla_z\}\,$. In film's interior
$\bm{\nabla}^2\widetilde{U}_n$ $=0$, therefore the potentials of
conductors have the form
\begin{equation}
\widetilde{U}_n(\bm{k},z)=\Phi _{n}(\bm{k})\exp\{|\bm{k}|(\sigma
_{n}z-D/2)\}\,, \label{eq:ff}
\end{equation}
where $\,\sigma _n=1$ ($-1$), if $n$-th conductor is placed above
(below) film, and form-factor $\Phi _n(\bm{k})$ describes
distribution of $n$-th current. In combination with
(\ref{eq:lmax}) and (\ref{eq:emf}) the latter formula implies the
relation between the impedances, from one hand, and values of the
potentials taken at film's surfaces, from another hand:
\begin{equation}
Z_{nm}=\frac{i\omega }{2\pi }\int |\bm{k}| \widetilde{U}_{sm}\left
(\omega ,\bm{k},\sigma _{n} \frac {D}{2}\right
)\widetilde{U}_n\left (-\bm{k}, \sigma _{n}\frac {D}{2}\right
)d\bm{k} \label{eq:surf}
\end{equation}
Here $\,d\bm{k}$ $\equiv $ $dk_xdk_y$ $/(2\pi)^2\,$ (in contrast
with $\widetilde{U}_n\,$, potentials $\widetilde{U}_{sn}$ are
frequency dependent). Thus, the potentials are taken at the
surface most close to receiving antenna ($n$-th conductor).

The Eq.\ref{eq:fr}, or equivalently,
\begin{equation}
(\bm{\nabla}\cdot \widehat{\mu }\bm{\nabla})
(\widetilde{U}_{sn}+\widetilde{U}_n)=0\,\,, \label{eq:peq}
\end{equation}
should be solved under standard boundary conditions \cite{ABP,LL}.
To write the answer, introduce the unit-length vectors
\[
\bm{\nu}\equiv \{k_x/|\bm{k}|,k_y/|\bm{k}|,0\}\,,\,\,\,
\,\,\overline{\bm{z}}\equiv\{0,0,1\}\,,
\]
and besides the matrix
\begin{equation}
M=\left[
\begin{array}{cc}
\mu _{\nu \nu } & \mu _{\nu z} \\
\mu _{z\nu } & \mu _{zz}
\end{array}
\right] \equiv \left[
\begin{array}{cc}
\bm{\nu}\cdot\widehat{\mu }\bm{\nu} & \bm{\nu}\cdot\widehat{\mu }
\overline{\bm{z}} \\
\overline{\bm{z}}\cdot\widehat{\mu }\bm{\nu} &
\overline{\bm{z}}\cdot\widehat{\mu }\overline{\bm{z}}
\end{array}
\right]    \label{M}
\end{equation} As usually, the solution is
composed by two exponents:
\begin{equation}
\widetilde{U}_{sn}+\widetilde{U}_n= \sum_{\pm }u_{n\pm }\exp
(q_{\pm }z)\,,\,\,\,\,\, q_{\pm }=\lambda _{\pm}|\bm{k}|\,,
\label{eq:u}
\end{equation}
\begin{equation}
\lambda _{\pm }\equiv \lambda _0\pm \Lambda \,,\,\,\,\,\,
\,\lambda _0=\frac{\mu _{\nu z}+
\mu _{z\nu }}{2i\mu _{zz}}\,,              \label{eq:lam0}
\end{equation}
\begin{equation}
\Lambda =\sqrt{ \frac{\mu _{\nu \nu }}{\mu _{zz}}-
\left( \frac{\mu _{\nu z}+\mu _{z\nu }}{2\mu _{zz}}
\right) ^{2} }                           \label{eq:lam}
\end{equation}
It should be emphasized, however, that in general case (at
arbitrary orientation of the vector $\bm{S}_0$) the exponents
$q_{\pm }\,$ are neither poorly imaginary nor poorly real but
complex, that is MSW is not standing in $Z$-direction.

If we took into account finite exchange radius $r_e\,$ and dealt
with the operator-valued polarizability matrix (\ref{eq:hi}), then
in place of (\ref{eq:u}) we would get a sum of at least six terms,
where transverse wave numbers of order of $|\bm{k}|$ (as
$q_{\pm}\,$ in (\ref{eq:u})) are more or less hybridized with real
or imaginary wave numbers of order of $\,\pi/r_e\,$. From the
point of view of our aims, such complication would be meaningless
accuracy. But it can be necessary when considering short SW or
small-scale details of long MSW on background of a nonuniform
domain structure. Most natural approach to these tasks is direct
analysis of the system of equations (\ref{eq:lin}) and
(\ref{eq:lmax}).

Consider the susceptibility matrix $M$. In standard spherical
coordinates let $\,\theta \,$ be the angle between $Z$-axis and
$\bm{S}_0$. In the $XY$-plane (film's plane), we introduce
quantities $\,\nu _{\Vert }\,$ and $\,\nu _{\bot }\,$ as cosine
and sine, respectively, of the angle (counted clockwise) between
projection of $\bm{S}_0$ onto this plane and the above defined
unit vector $\,\bm{\nu}\,$ lying in it. Besides, consider the
plane $\,\bm{a}_1\bm{a}_2$ perpendicular to $\bm{S}_0$ and, in
this plane, define $\,\psi\,$ being the angle between plane
$Z\bm{S}_0$ and vector $\,\bm{a}_1\,$ (definition of vectors
$\,\bm{a}_1\,$ and $\,\bm{a}_2\,$ was done at the end of Sec.3).
Further, introduce the quantities $\,\,A_{\pm}\equiv $ $(A_2\pm $
$A_1)/2\,$. At last, let $\Omega\,$ be nominator of the
polarizability matrix (\ref{eq:mhi}),
\[
\Omega =(\overline{W}_0+A_{1}+A_{2}) \widehat{\Pi
}-\widehat{A}_{\bot }+i\widetilde{\omega } \widehat{R}
\]
In these designations
\begin{equation}
\Omega _{zz}=(\overline{W}_0+A_{+}+A_{-}\cos \,2\psi ) \sin
^2\theta \,\,,              \label{eq:ozz}
\end{equation}
\begin{equation}
\Omega _{\nu \nu }=(\overline{W}_0+A_{+})
(\nu _{\bot }^{2}+
\nu _{\Vert }^{2}\cos ^{2}\theta )+    \label{eq:onn}
\end{equation}
\[
+A_{-}\{(\nu _{\Vert }^{2}\cos ^{2}\theta -\nu _{\bot }^{2}) \cos
\,2\psi -2\nu _{\Vert }\nu _{\bot }\sin \,2\psi \cos \,\theta
\}\,\,,
\]
\begin{equation}
\Omega _{z\nu ,\,\nu z}=\Omega _{\times } \mp i\omega \nu _{\bot
}\sin \,\theta \,\,,    \label{eq:onz}
\end{equation}
\begin{equation}
\Omega _{\times }\equiv \sin \,\theta \,\{A_{-}\nu _{\bot } \sin
\,2\psi\, -  \label{eq:ocr}
\end{equation}
\[
-\,\nu _{\Vert }(\overline{W}_{0}+A_{+} +A_{-}\cos \,2\psi )\cos
\,\theta \}
\]
These formulas make it evident that effects of anisotropy are
determined by $A_{-}\,$, while $A_{+}\,$ merely redefines the
magnitude of static internal field, $W_0$.

Then, for a given in-plane orientation of the wave,
$\,\bm{\nu}\,$, introduce characteristic frequencies by the
following expressions:
\begin{equation}
\omega _0^2\equiv (\overline{W}_0+A_{+})^2 -A_{-}^2\,\,,
\label{eq:w0}
\end{equation}
\begin{equation}
\omega _u^2\equiv \omega _0^2+ 4\pi \Omega _{zz}\,\,,\label{eq:wu}
\end{equation}
\begin{equation}
\omega _{1,2}^{2}\equiv \omega _0^2+ 2\pi (\Omega _{zz}+\Omega
_{\nu \nu })\,\mp   \label{eq:w12}
\end{equation}
\[
\mp\, 2\pi \sqrt{(\Omega _{zz}+ \Omega _{\nu \nu })^{2}-(2\nu
_{\bot } \sin \,\theta )^{2}\omega _0^2}\,\,\,,
\]
\begin{equation}
\omega _{3}^{2}\equiv \{\omega _{1}^{2}+\omega _{2}^{2}+(4\pi \nu
_{\bot }\sin \,\theta )^{2}\}/2  \label{eq:w3}
\end{equation}
The frequency $\omega _u\,$ which is independent on the in-plane
wave vector $\bm{k}\,$ is the uniform precession frequency. In
terms of these
 frequencies,
\begin{equation}
\mu _{zz}=\frac{\omega _u^2-\omega ^2 }
{\omega _0^2-\omega ^2}\,,\,\,\,\,
\mu _{z\nu }-\mu _{\nu z}=-\frac{8\pi i\omega
\nu _{\bot }\sin \,\theta }
{\omega _0^2-\omega ^2}\,,                   \label{eq:mus}
\end{equation}
\begin{equation}
\lambda _0=\frac{4\pi i\Omega _{\times }}
{\omega ^2-\omega _u^2}\,,\,\,\,\,\,\,
\Lambda =\frac{\sqrt{(\omega _1^2-\omega ^2)
(\omega _2^2-\omega ^2)}}{\omega ^2-\omega _u^2}  \label{eq:lams}
\end{equation}
Besides, below we will need in the determinant
\begin{equation}
\Delta \equiv \det \,M\,= \frac{2\omega _3^2-\omega _0^2
-\omega ^2}{\omega _0^2-\omega ^2}            \label{eq:del}
\end{equation}
It appears that $\Delta \,$ being quadratic function of the matrix
elements of $M\,$ and $\widehat{\chi }\,$ nevertheless always has
simple pole only.

We omit trivial but tremendous evaluation of the surface
potentials which appear in Eq.\ref{eq:surf}. The result, for the
surface closest to a given antenna, looks as
\begin{equation}
\widetilde{U}_{sn}(\omega ,\bm{k},\sigma _{n}D/2) =\Phi
_{n}(\bm{k})P(\omega ,\bm{k})\,\,,           \label{eq:bp}
\end{equation}
\begin{widetext}
\begin{equation}
P(\omega ,\bm{k})\equiv \frac{1-\Delta -i(\mu _{z\nu }-\mu _{\nu
z})}{1+\Delta +2\mu _{zz}\Lambda \coth (\Lambda |\bm{k}|D)}
=\frac{\omega _0^2-\omega _3^2 -4\pi \omega \nu _{\bot }\sin
\,\theta }{G(\omega ,\bm{k})}\,\,, \label{eq:prop}
\end{equation}
\end{widetext}
where denominator in the latter expression is given by
\begin{equation}
G(\omega ,\bm{k})\equiv \omega _3^2 -\omega ^2 +(\omega
_u^2-\omega ^2)\Lambda \coth \,(\Lambda |\bm{k}|D) \label{eq:G}
\end{equation}
For brevity, we do mark dependencies of the factors $\Delta $ and
$\Lambda $ on $\omega\,$ and $\bm{k}\,$ as well as dependencies of
$\omega _{1,2,3}\,$ on  $\bm{k}\,$ (or, to be precise, on
direction of the in-plane wave vector $\bm{k}$). Combining these
formulas and Eq.\ref{eq:surf}, for mutual impedance of two
antennas located on one and the same side from film, we obtain:
\begin{equation}
Z_{nm}=\frac{i\omega }{2\pi }\int |k|\Phi _{n}(-k)
\Phi _{m}(k)P(\omega ,k)dk               \label{eq:imp}
\end{equation}

The latter formulas present main results of the paper and, as far
as we know, can give useful addition to results of Damon and
Eshbach \cite{de} and other authors (see Sec.1). Function
$P(\omega ,k)\,$ is the required propagator of linear (weak)
magnetostatic excitations from one antenna to another. At the same
time, it contains complete information about spectrum of MSW. The
condition that its denominator turns into zero, $G(\omega ,k)$
$=0$ (in absence of dissipation, at  $\gamma =0$), yields a set of
dispersion laws for all possible types of MSW. This will be the
subject of separate work.

\section{Mutual impedance of wire antennas}
To be more concrete, consider relatively simple but practically
interesting case of straight-line wire antennas which have round
cross-sections and are parallel one to another and to film's
surface. Besides, let they be oriented along $Y$-axis and located
at $X$-positions $x_{n}\,$, on the same side from the film and at
distances $\rho _{n}\,$ from its closest surface. In this
situation
\[
\Phi _n(k)=(4\pi ^2/ick_x) \exp (-|k_x|\rho _n-ik_x x_n)\delta
(k_y)\,\,
\]
(here $\,c\,$ is speed of light), and Eq.\ref{eq:imp} concretizes
to
\begin{widetext}
\begin{equation}
\frac {Z_{nm}[\text{Ohm}]}{w[\text{cm}]f[\text{GHz}]} =4\pi
i\int_{0}^{\infty } \exp \{-q(\rho _n+\rho _m)\}\cdot
\frac{(1-\Delta )\cos (qx)+(\mu _{z\nu }- \mu _{\nu z})\sin
(qx)}{1+\Delta +2\mu _{zz}\Lambda \coth (\Lambda qD)} \cdot \frac
{dq}{q}   \label{DW}
\end{equation}
\end{widetext}
Here on the left $\,w\,$ is the film's width (formally infinite)
measured in centimeters along antennas (i.e in $Y$-direction),
$\,f\,$ is the frequency expressed in GHz, while on the right-hand
side $\,x\equiv $ $x_n-x_m\,$, and the integral is taken over
$\,q\equiv k_x\,$ at $\,k_y\rightarrow 0$. The latter means that
matrix elements of the magnetic susceptibility matrix $M$ and the
functions $\Lambda $ and $\Delta $ are calculated at $\,\nu =$
$\{1,0,0\}$.

Note that the dimensionless circular frequency $\,\omega \,$ which
enters all these functions is connected with factual frequency
$\,f\,$ expressed in GHz by the relation (see Sec.2)
\[
\omega =f/f_0\,\,\,,\,\,\,f_0\equiv (2\pi\tau_0)^{-1}=gM_s
\]

The impedance (\ref{DW}) possesses evident asymmetry with respect
to sign of $\,x\,$, if $\,(\mu _{z\nu }-\mu _{\nu z})$ $\neq 0$.
This is one more example of the non-reciprocity inherent to wave
propagation in presence of static magnetization.

One can see also that the impedance is a function of the
dimensionless ratios $D/\rho_n\,$ and $x/D\,$ only. Clearly, this
is the consequence of specific scale invariance of the dipole
interaction which is sensible to shape of the sample but not its
size.

In obvious way, one can generalize Eq.\ref{DW} for many-element
antennas each consisting of several (mutually parallel) wires with
alternated signs of current in them. It should be noted also that
all the formulas (\ref{eq:ozz})-(\ref{eq:del}) will serve for
analytical calculations (e.g. in next our publication), but if
using computer it is sufficient to numerically calculate matrix
(\ref{M}) and then factors (\ref{eq:lam}) and (\ref{eq:del}) only
(by this reason we expressed the impedance in terms of these
quantities). In general, of course, first of all one must find the
static magnetization vector, $\bm{S}_0$, but this is also not a
hard task for computer.

For the case when two parallel wires are situated on the opposite
parties from the film, evaluation of corresponding boundary
potentials yields (in the same units):
\begin{widetext}
\begin{equation}
\frac{Z_{nm}}{wf\,}=2\pi i\int_{-\infty }^{\infty }e^{-|q|(\rho
_{n}+\rho _{m})+iqx}\left\{ e^{-|q|D}- \frac{2\mu _{zz}\Lambda
\exp (\lambda _{0}|q|D)}{(1+\Delta )\sinh (\Lambda |q|D)+ 2\mu
_{zz}\Lambda \cosh (\Lambda |q|D)} \right\} \frac{dq}{|q|}
                \label{os}
\end{equation}
\end{widetext}
Of course, here in the integrand $\,\bm{\nu} =$
$\{$sign$(q),0,0\}$, and $\lambda _0\,$ is defined in
(\ref{eq:lams}).

For simple example, let us evaluate self-impedance, $Z_{11}$, of
straight wire antenna (to be precise, the contribution to full
self-impedance stipulated by the film), in the special case when
bias magnetic field vector, $\bm{H}_0$, lies in the film's plane.
For concreteness, let it be oriented along $Y$-axis. Besides, we
assume that characteristic magnetic field of anisotropy is small
in comparison with $|\bm{H}_0|+4\pi M_s$, which allows to neglect
effects of anisotropy. At last, if we direct the antenna in
parallel to $\bm{H}_0$ then the impedance must be caused primarily
by the so-called surface MSW discovered by Damon and Eshbach
\cite{de}. Under above formulated conditions, in $\bm{k}$-plane
these waves occupy the sector $\,\left|k_y/k_x\right|$ $<$ $\sqrt
{4\pi M_s/|\bm{H}_0|}\,$. But, naturally, sufficiently
(infinitely) long antenna excites mainly the waves with
$|k_y/k_x|\rightarrow 0$ which run perpendicularly to the field.
Apparently, the latter case is the only case when the dispersion
law of MSW (Damon-Eshbach waves with $\,\bm{k}\perp \bm{H}_0$, or
DE-waves) can be written in the evident analytical form \cite{de}:
\begin{equation}
\omega _{DE}(\bm{k})=\sqrt{|H_0|(|H_0|+4\pi )+ 4\pi ^2[1-\exp
(-2D|\bm{k}|)]}           \label{eq:DE}
\end{equation}
Here field and frequency are expressed in the dimensionless units
introduced in Sec.2.

In this case there is a good analytical approximation for the
integral (\ref{DW}) which yields
\begin{widetext}
\begin{equation}
\frac {R_{11}[\text{Ohm}]}{w[\text{cm}]f[\text{GHz}]}\approx \frac
{4\pi \omega _t\,X}{1-X^2} \left (\frac {1-X}{1+X} \right )^{\rho
/2D} \left[\ln\frac {1+X}{1-X}\right]^{-1} \,\,\,, \label{eq:R}
\end{equation}
\[
X\equiv \frac {\omega ^2-\omega _u^2}{4\pi ^2}\,\,,\,\,
\,\,\,\omega =\frac {f}{f_0}\,\,,\,\,\, \omega _u=
\sqrt{|\bm{H}_0|(|\bm{H}_0|+4\pi )}\,\,,\,\,\, \omega _t=
|\bm{H}_0|+2\pi
\]
\end{widetext}
In this formula, $R_{11}=$\,Re$\,Z_{11}\,$ is the film-induced
contribution to resistance of the antenna ($Z=$ $R-$ $2\pi ifL$);
magnetic field is dimensionless, i.e. expressed in units of $M_s$;
the frequency belongs to the interval $\,f_0\omega _u$ $<f$
$<f_0\omega _t\,\,$; $\,\omega _u$ is dimensionless frequency of
uniform precession and at the same time lower bound of spectrum of
the Damon-Eshbach waves, while $\,\omega _t\,$ is upper bound of
this spectrum (and spectrum of MSW at all). Outside this frequency
interval, there are no waves perpendicular to the antenna, and
hence $\,R_{11}\,$ turns into zero (or, to be more precise,
becomes comparatively small).

Let us notice that $X\rightarrow 1\,$ when $\,f\rightarrow $
$f_0\omega _t\,$, therefore under condition $\,\rho /2D$ $<1\,$
the resistance (\ref{eq:R}) tends to infinity at upper edge of
spectrum of the DE-waves. The matter is that here the group
velocity of DE-waves, $v_g\,$, turns into zero, hence density of
states (DE-wave modes) tends to infinity.

Under the opposite condition, $\,\rho /2D$ $>1\,$, this effect is
canceled by sufficient weakness of excitation of short DE-waves.
The presence in (\ref{eq:R}) of the exponent which depends on
geometric parameters of the system is eventually consequence of
scale invariance of the dipole interaction.

What is interesting, in the lower part of spectrum of DE-waves the
resistance $R_{11}$ is almost independent on the film's thickness,
$D$, although seemingly the e.m.f. and thus the resistance must be
proportional to amount of magnetic moments (spins) under
excitation and thus to $D\,$. The matter is that the energy
outflow from the antenna, $p\,$, is proportional to the group
velocity, $\,v_g\,$, of the DE-waves under excitation: $p\propto $
$Dv_g$ $|\bm{S}_{\bot }|^2\,$ (where $\bm{S}_{\bot }\,$ represents
magnitude of spin precession). From the other hand, we can write
$\,p\propto $ $\varepsilon _1^2$ $/R_{11}\,$, while $\,\varepsilon
_1 \propto$ $D|\bm{S}_{\bot }|$. Three above relations result in
$R_{11}\propto $ $D/v_g\,$. But, as it follows from (\ref{eq:DE}),
group velocity of long DE-waves is proportional to the thickness,
$\,v_g\propto D$. This is the reason for the indifference of
$\,R_{11}(\omega\rightarrow$ $\omega _u)\,$ with respect to $D$.

A simple analytical estimate for the inductance, $L_{11}$
$=-\,$Im$\,Z_{11}/(2\pi f)\,$, can be deduced at $\rho /D$ $\sim
1\,$ only, and then
\begin{equation}
\frac{L_{11}[\text{nH}]}{w[\text{cm}]} \approx -(2+|H_0|/\pi )
\exp (-X) \text{Ei}(X)            \label{eq:L}
\end{equation}
($\,\text{Ei}\,$ is the integral exponent function). Clearly,
$L_{11}\,$ can be both positive and negative.

\section{Conclusion}
In brief, we found (i) propagator (\ref{eq:prop}) of magnetostatic
waves (MSW) running in infinite ferromagnetic film from one
antenna to another and (ii) linear (small-amplitude) mutual
impedance of the antennas, under arbitrary orientation of uniform
static magnetization of the film and arbitrary magnetic
anisotropy.

Additionally, the equation $G(\omega ,\bm{k})=0$ with $G$ being
the denominator (\ref{eq:G}) in (\ref{eq:prop}) determines
dispersion laws for various linear eigenmodes of MSW thus allowing
generalizations of classical results obtained in \cite{de}.

To conclude, let us touch the applicability of formulas, obtained
for continuous spectrum of MSW in infinite film, to real
finite-size films where MSW spectrum is discrete. When in-plane
dimensions of a film decrease then characteristic frequency
separation between neighbouring MSW modes increases, but, at the
same time, selection of the modes by any simple (for instance,
straight-line wire) antenna becomes more and more worsened. As the
result, the formulas derived for infinite film can give good
estimate for impedances of antennas interacting with real films.

This expectation was confirmed by comparison between the
analytical estimates and measurements of impedances induced by
millimeter-size ferrite films as well as by results of their
numeric simulations. Moreover, numeric simulations of the torque
equation (\ref{eq:LL}) with dipole-dipole interactions between
spins show that spatial-temporal patterns of spin precession even
in rather small films (with length to thickness ratio $\sim 30$
$\div 100$) and even at essentially non-linear regimes possess
clear imprints of qualitative and quantitative characteristics
inherent to linear MSW modes in infinite system.




\end{document}